\documentclass[12pt]{ussci}
\pdfoutput=1
\usepackage{enumitem}
\setlist{noitemsep}

\usepackage{xmpincl}
\includexmp{CC_Attribution_4.0_International}

\hypersetup{
    pdftitle={UConnRCMPy: Python-based data analysis for rapid compression machines},%
    pdfauthor={Bryan W. Weber},%
}

\usepackage{siunitx}
\usepackage{graphicx}
\usepackage{mathtools}
\usepackage{microtype}

\usepackage{textcomp}
\DeclareUnicodeCharacter{3BC}{\textmu}
\DeclareUnicodeCharacter{39C}{\textmu}

\usepackage[capitalize]{cleveref}
\newcommand\papertopic{Reaction Kinetics}
\DeclareSIUnit\torr{torr}

\title{ UConnRCMPy: Python-based data analysis for rapid compression machines }

\author[1*]{Bryan W.\ Weber}
\author[1]{Chih-Jen Sung}
\affil[1]{Department of Mechanical Engineering, University of Connecticut, Storrs,
CT, USA}
\affil[*]{Corresponding Author: \email{bryan.weber@uconn.edu}}

\begin{document}
\maketitle

\begin{abstract} 
    The ignition delay of a fuel/air mixture is an important quantity in
    designing combustion devices, and these data are also used to validate
    chemical kinetic models for combustion. One of the typical experimental
    devices used to measure the ignition delay is called a Rapid Compression
    Machine (RCM). This paper presents UConnRCMPy, an open-source Python package
    to process experimental data from the RCM at the University of Connecticut.
    Given an experimental measurement, UConnRCMPy computes the thermodynamic
    conditions in the reaction chamber of the RCM during an experiment along
    with the ignition delay. UConnRCMPy implements an extensible framework, so
    that alternative experimental data formats can be incorporated easily. In
    this way, UConnRCMPy improves the consistency of RCM data processing and
    enables the community to reproduce data analysis procedures.
\end{abstract}


\section{Introduction}\label{introduction}

In recent years, there has been a surge in interest in ensuring that research
outputs are reproducible across time and personnel~\autocite{NatureEds2016}.
Recognizing that the code used to process experimental data is an important part
of the chain from observation to result and publication, this paper presents the
design and operation of a software package to process the pressure data
collected from Rapid Compression Machines (RCMs). Our package, called
UConnRCMPy~\autocite{uconnrcmpy}, is designed to analyze the data acquired from
the RCM at the University of Connecticut (UConn). Despite the initial focus on
data from the UConn RCM, the package is designed to be extensible so that it can
be used for data in different formats while providing a consistent interface to
the user. Thus, UConnRCMPy offers all of the features required to process
standard RCM data including:

\begin{itemize}
\item
  Filtering and smoothing the raw voltage output generated by the pressure
  transducer
\item
  Converting the voltage trace into a pressure trace using settings
  recorded from the RCM
\item
  Processing the pressure trace to determine parameters of interest in
  reporting the experiments, including the ignition delay and
  machine-specific effects on the experiment
\item
  Conducting simulations utilizing the experimental information to
  calculate the temperature at the end of compression (EOC)
\end{itemize}

Previous software used to analyze RCM data has generally been undocumented and
untested code specific to the researcher conducting the experiments. Moreover,
the software typically used to estimate the temperature in the experiments is
difficult to integrate with the data processing code. To the best of the
authors' knowledge, UConnRCMPy is the first package for analysis of standard RCM
data to be presented in detail in the literature, and it tightly integrates the
temperature estimation routine into the workflow, reducing errors and
inefficiencies.

\section{RCM Signal Processing Procedure}\label{rcm-signal-processing-procedure}

The RCMs at the University of Connecticut have been described extensively
elsewhere~\autocite{Das2012,Mittal2007a}, and interested readers are referred to
those papers for further details. The primary diagnostic on the RCM is the
reaction chamber pressure during and after the compression process, measured by
a dynamic pressure transducer. The pressure trace is processed to determine the
quantities of interest, including the pressure and temperature at the EOC,
\(P_C\) and \(T_C\) respectively, and the ignition delay, \(\tau\). These values
depend on the pressure and temperature prior to the start of compression
(\(P_0\) and \(T_0\), respectively), in addition to the composition of the
reactant mixture and the overall compression ratio of the RCM. A single
compression-delay-ignition sequence is referred to as an experiment or a run and
a set of experiments at a given \(P_C\) and mixture composition is referred to
as a condition.

\begin{figure}[htbp]
    \begin{minipage}[t]{0.48\textwidth}
        \centering
        \includegraphics[width=\linewidth]{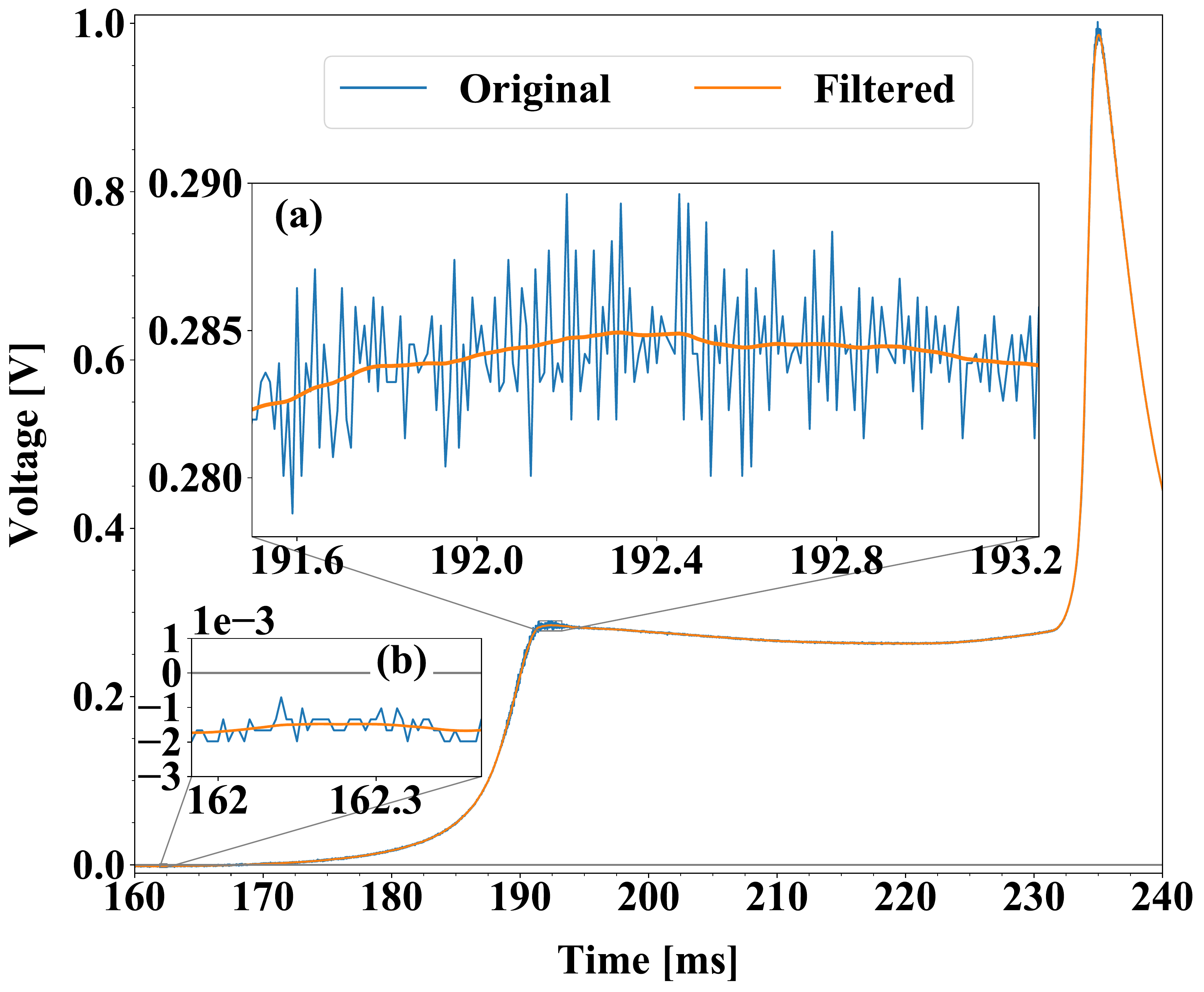}
        \caption{Raw voltage trace and the voltage trace after filtering from a typical
        RCM experiment. (a): Close up of the
        time around the EOC (b): Close up of the time before the start of
        compression}
        \label{fig:raw-voltage}
    \end{minipage}\hfill%
    \begin{minipage}[t]{0.48\textwidth}
        \centering
        \includegraphics[width=\linewidth]{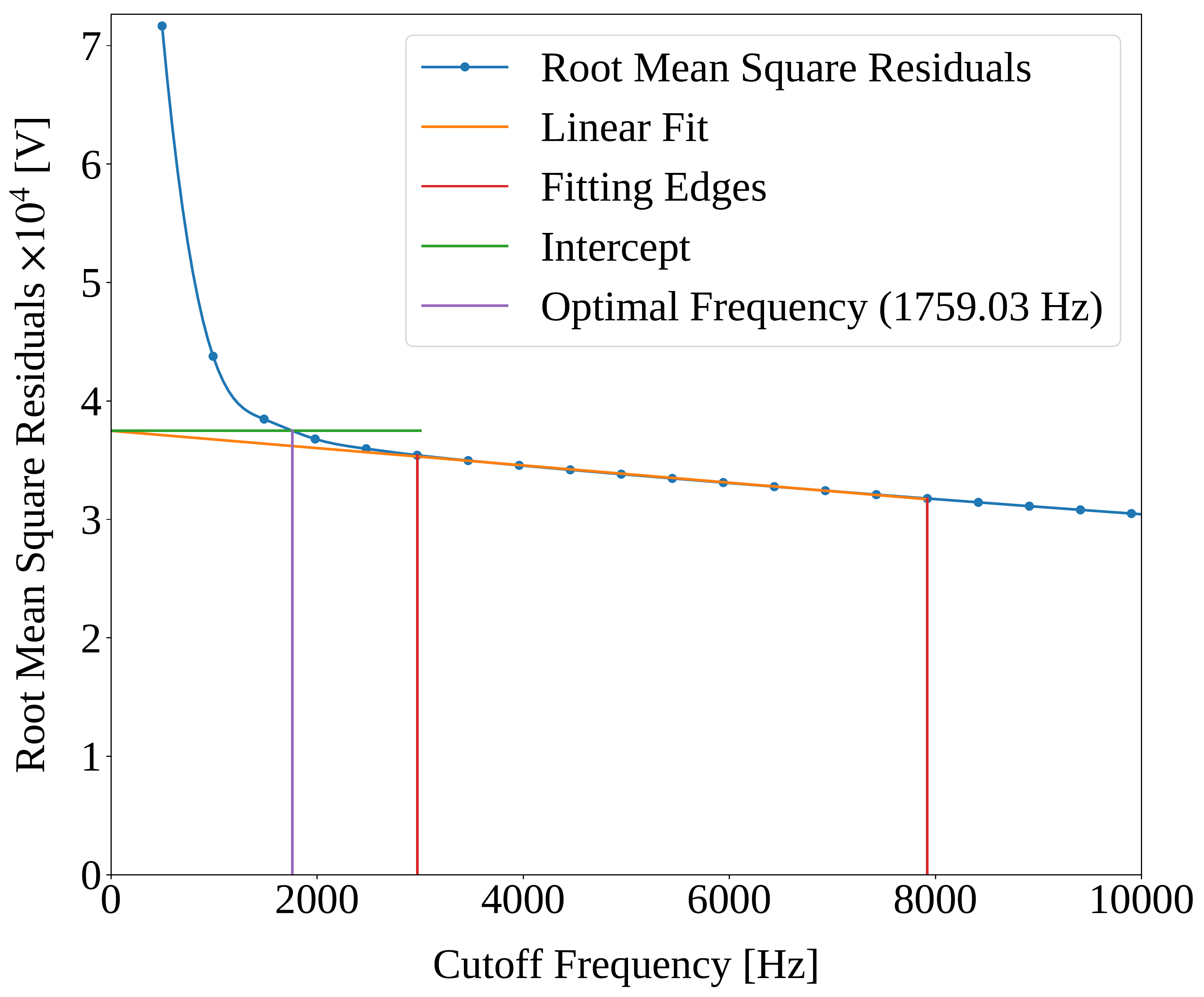}
        \caption{Root mean square residuals of the filtered signal compared to
        the original signal for a range of filter cutoff frequencies. The
        Nyquist frequncy for this case is \SI{50}{\kHz}. The left ``Fitting
        Edge'' is at the first frequency used for fitting greater than
        one-twentieth the Nyquist frequency}
        \label{fig:residuals}
    \end{minipage}
\end{figure}

The dynamic pressure transducer outputs a charge signal that is converted to a
voltage signal by a charge amplifier with a nominal output of \SI{0}{\V}. In
addition, the output range of \SIrange{0}{10}{\V} is set by the operator to
correspond to a particular pressure range by setting a ``scale factor.'' The
voltage output from the charge amplifier is digitized by a hardware data
acquisition system and recorded into a plain text file by a LabView Virtual
Instrument. \Cref{fig:raw-voltage} shows a typical voltage trace measured from
the RCM at UConn and demonstrates the typical noise in the signal, which
requires filtering and further processing to produce a useful pressure trace.

In the current version of UConnRCMPy~\autocite{uconnrcmpy}, the voltage is
filtered using a first-order Butterworth filter. The cutoff frequency of the
filter is chosen automatically by a procedure described in the work of
\textcite{Yu1999,Duarte2014}. Briefly, this procedure applies low-pass filters
of varying cutoff frequencies to the signal and calculates the root mean square
residual between the filtered signal and the original signal.
\Cref{fig:residuals} shows a typical plot of the residuals versus the cutoff
frequency and demonstrates that the residuals are nearly linear for a range of
cutoff frequencies. This range tends to start near one-twentieth the Nyquist
frequency, as demonstrated by the left-most line labeled ``Fitting Edges'' on
\cref{fig:residuals}. To determine the right edge of the linear region, a series
of linear regressions of the residuals are performed. The $y$-intercept of the
regression with the highest coefficient of determination is used to choose the
optimal cutoff frequency. The right-most line labeled ``Fitting Edges'' in
\cref{fig:residuals} demonstrates a case where the end point set at \num{0.15}
times the Nyquist frequency produces the best fit. The optimal cutoff frequency
is chosen as the frequency at the intersection of the $y$-intercept and the
residuals curve.

After filtering, the voltage trace is converted to a pressure trace by
correcting for the offset from the nominal initial volage of \SI{0}{\V} apparent
in \mbox{\cref{fig:raw-voltage}b}, multiplying the voltage by the scale factor
from the charge amplifier, and adding the initial pressure \(P_0\). The result
is a vector of time-varying pressure values that must be further processed to
determine the time of the EOC and the ignition delay.

Once the pressure trace has been constructed, \(T_C\), \(P_C\), and \(\tau\) can
be calculated. In the current version of UConnRCMPy~\autocite{uconnrcmpy}, the
time of the EOC is determined by finding the local maximum of the pressure prior
to ignition. Then, the ignition delay is determined as the time difference
between the EOC and the point of ignition, where the point of ignition is
defined as the inflection point in the pressure trace due to ignition. The
inflection point is found by the maximum of the first derivative of the pressure
with respect to time. In the current version of
UConnRCMPy~\autocite{uconnrcmpy}, the first derivative of the experimental
pressure trace is computed by a second-order forward differencing method. The
derivative is then smoothed by a moving average algorithm with a width of 151
points. This value for the moving average window was chosen empirically.

\begin{figure}[htbp]
    \centering
    \includegraphics[width=0.6\textwidth]{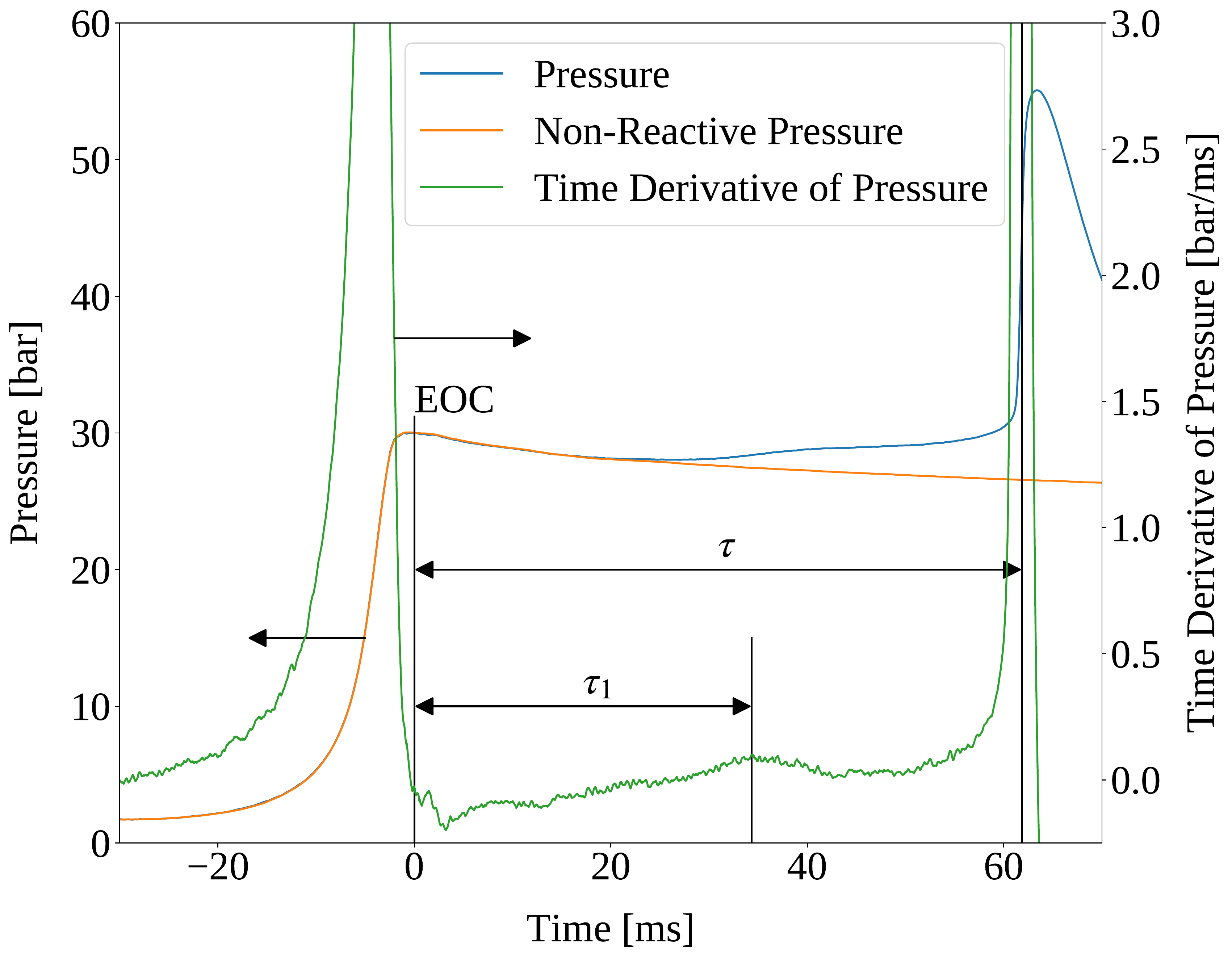}
    \caption{Illustration of the definition of the ignition delay in a
    two-stage ignition case}
    \label{fig:ign-delay-def}
\end{figure}

For some conditions, the reactants may undergo two distinct stages of ignition.
These cases can be distinguished by a pair of peaks in the first time derivative
of the pressure. For some two-stage ignition cases, the first-stage pressure
rise, and consequently the peak in the derivative, are relatively weak, making
it hard to distinguish the peak due to ignition from the background noise. This
is currently the area requiring the most manual intervention, and one area where
significant improvements can be made by refining the differentiation and
filtering algorithms. An experiment that shows two clear peaks in the derivative
is shown in \cref{fig:ign-delay-def} to demonstrate the definitions of the
ignition delays.

The final parameter of interest presently is the EOC temperature, \(T_C\). This
temperature is often used as the reference temperature when reporting ignition
delays. In general, it is difficult to measure the temperature as a function of
time in the reaction chamber of the RCM, so methods to estimate the temperature
from the pressure trace are used. The detailed procedure used in UConnRCMPy is
described in the work of \textcite{Dames2016}, and an overview is given here.

In general, the temperature in the RCM reaction chamber as a function of time
can be found by integrating the first law of thermodynamics for an ideal gas:
\begin{equation} \label{eq:first-law}
    c_v \frac{dT}{dt} = -P \frac{dv}{dt} - \sum_k u_k \frac{d Y_k}{dt}
\end{equation}
where \(c_v\) is the specific heat at constant volume of the mixture, \(v\) is
the specific volume, \(u_k\) and \(Y_k\) are the specific internal energy and
mass fraction of the species \(k\), and \(t\) is time. In UConnRCMPy,
\cref{eq:first-law} is integrated by Cantera~\autocite{cantera}.

Integrating \cref{eq:first-law} requires knowledge of the volume of the reaction
chamber as a function of time. To calculate the volume as a function of time, it
is assumed that there is a core of gas in the reaction chamber that undergoes an
isentropic, constant composition compression~\autocite{Lee1998}. The initial
entropy of the gas mixture is calculated using Cantera~\autocite{cantera}.
Subsequently, the state of the mixture is fixed by using the entropy and
measured pressure; from this information, the volume is calculated. The initial
volume is arbitrarily taken to be \(V_0=\SI{1.0}{\m\cubed}\). The initial volume
used in constructing the volume trace is arbitrary provided that the same value
is used for the initial volume in the simulations described below. However,
extensive quantities such as the total heat release during ignition cannot be
compared to experimental values.

Two simulations can be triggered by the user that solve \cref{eq:first-law}. In
the first, the multiplier for all the reaction rates is set to zero, to simulate
a constant composition (non-reactive) process. In the second, the reactions are
allowed to proceed as normal. Only the non-reactive simulation is necessary to
determine \(T_C\), which is defined as the simulated temperature at the EOC
time.

When a reactive simulation is conducted, the user must compare the
temperature traces from the two simulations to verify that the inclusion
of the reactions does not change \(T_C\), validating the assumption of
adiabatic, constant composition compression. Although including
reactions during the compression stroke does not affect the value of
\(T_C\), it does allow for the buildup of a small pool of radicals that
can affect processes after the EOC~\autocite{Mittal2008}. Thus, it is
critical to include reactions during the compression stroke when
conducting simulations to compare a kinetic model to experimental
results.

As can be seen in \cref{fig:ign-delay-def}, the pressure
decreases after the EOC due to heat transfer from the higher temperature
reactants to the reaction chamber walls. This process is specific to the
machine that carried out the experiments, and to the conditions under
which the experiment was conducted. To include the effect of this heat
transfer into simulations, a non-reactive experiment is conducted, where
\(\text{O}_2\) in the oxidizer is replaced with \(\text{N}_2\).

To apply the effect of the post-compression heat loss into the simulations, the
reaction chamber is modeled as undergoing an isentropic volume expansion after
EOC, and the same procedure is used as in the computation of \(T_C\) to compute
a volume trace for the post-EOC time. The only difference is that the
non-reactive pressure trace is used after the EOC instead of the reactive
pressure trace. This procedure has been validated experimentally by measuring
the temperature in the reaction chamber during and after the compression stroke.
The temperature of the reactants was found to be within $\pm\sim $\SI{5}{\K} of
the simulated temperature~\autocite{Das2012a,Uddi2012}.

\section{Implementation and Usage of UConnRCMPy}\label{implementation-and-usage-of-uconnrcmpy}

UConnRCMPy is constructed in a hierarchical manner. The main user interface to
UConnRCMPy is through the \pybox{Condition} class, the highest level of data
representation. The \pybox{Condition} class contains all of the information
pertaining to the experiments at a given condition. The intended use of this
class is in an interactive Python interpreter (the authors prefer the Jupyter
Notebook with an IPython kernel~\autocite{Perez2007}). First, the user creates
an instance of the \pybox{Condition} class.

\begin{pythonbox}
from uconnrcmpy import Condition

cond_00_02 = Condition(cti_file='./species.cti')
\end{pythonbox}

The \pybox{cti_file} argument to \pybox{Condition} must point to a file in the
CTI format that contains the thermodynamic and reaction information for the
species in the mixture. The experiments in the following example were conducted
with mixtures of propane, oxygen, and nitrogen~\autocite{Dames2016}. The CTI
file necessary to run this example can be found in the Supplementary Material of
the work by \textcite{Dames2016}. Then, the composition of the mixture under
consideration must be added to the \pybox{initial_state} parameter of the
\pybox{ideal_gas()} function in the CTI file:

\begin{pythonbox}
ideal_gas(
    name='gas', elements=..., species=..., reactions='all',
    initial_state=state(mole_fractions='C3H8:0.0403,O2:0.1008,N2:0.8589'))
\end{pythonbox}

Ellipses indicate input that was truncated to save space; the truncated input is
present in the file available with the work of \textcite{Dames2016}. The
\pybox{mole_fractions} must be set to the appropriate values. The condition in
this example is for a fuel rich mixture, with a target \(P_C\) of \SI{30}{\bar}.

After initializing the \pybox{Condition}, the user conducts a reactive
experiment with the RCM and adds the experiment to the \pybox{Condition} using
the \pybox{add_experiment()} method. As each experiment is processed by
UConnRCMPy, the information from that run is added to the system clipboard for
pasting into some spreadsheet software. In the current version, the information
copied is the time of day of the experiment, the initial pressure, the initial
temperature, the pressure at the EOC, the overall and first stage ignition
delays, an estimate of the EOC temperature, some information about the
compression ratio of the reactor, and the filter frequnecy used.

\begin{pythonbox}
# Conduct reactive experiment #1 on the RCM
cond_00_02.add_experiment('00_in_02_mm_373K-1282t-100x-19-Jul-15-1633.txt')
# ... conduct and add other reactive experiments
\end{pythonbox}

In general, for a given condition, the user will conduct and process all of the
reactive experiments before conducting any non-reactive experiments. Then, the
user chooses one of the reactive experiments as the reference experiment for the
condition (i.e., the one whose ignition delay(s) and \(T_C\) are reported) by
inspection of the data in the spreadsheet. The reference experiment is defined
as the experimental run whose overall ignition delay is closest to the mean
overall ignition delay among the experiments at a given condition. To select the
reference experiment, the user sets the \pybox{reactive_file} attribute of the
\pybox{Condition} instance. For this case, the reference experiment is the run
that took place at 16:33:

\begin{pythonbox}
cond_00_02.reactive_file = '00_in_02_mm_373K-1282t-100x-19-Jul-15-1633.txt'
\end{pythonbox}

Once the reference reactive experiment is selected, the user conducts
experiments at the same initial pressure and temperature conditions, but with a
non-reactive mixture. The user adds non-reactive experiments to the
\pybox{Condition} by the same \pybox{add_experiment()} method and UConnRCMPy
automatically determines whether the experiment is reactive or non-reactive. If
the user does not specify the \pybox{reactive_file} attribute, they are
prompted for the file name when the first non-reactive case is added.

\begin{pythonbox}
# Conduct non-reactive experiment #1 on the RCM
cond_00_02.add_experiment('NR_00_in_02_mm_373K-1278t-100x-19-Jul-15-1652.txt')
\end{pythonbox}

UConnRCMPy determines that this is a non-reactive experiment and generates a new
figure that compares the current non-reactive case with the reference reactive
case. For this particular example, the pressure traces are shown in
\cref{fig:ign-delay-def}. In this case, the non-reactive pressure agrees very
well with the reactive pressure and no further experiments are necessary; in
principle, any number of non-reactive experiments can be conducted and added to
the figure for comparison. Since there is good agreement between the
non-reactive and reactive pressure traces, the user sets the
\pybox{nonreactive_file} attribute of the \pybox{Condition} instance.

\begin{pythonbox}
cond_00_02.nonreactive_file='NR_00_in_02_mm_373K-1278t-100x-19-Jul-15-1652.txt'
\end{pythonbox}

Once the non-reactive case is chosen, the \pybox{create_volume_trace()} method
can be run. This method requires three attributes to be set on the
\pybox{Condition} instance: \pybox{nonreactive_end_time} which controls the
end time for volume trace generation, \pybox{reactive_end_time} which controls
the length of the pressure trace stored in the output file, and
\pybox{reactive_compression_time} which is the length of the compression
stroke. All of the values must be supplied in units of milliseconds.

\begin{pythonbox}
cond_00_02.nonreactive_end_time = 400
cond_00_02.reactive_end_time = 80
cond_00_02.reactive_compression_time = 36
\end{pythonbox}

After generating the volume trace, \pybox{create_volume_trace()} writes the
\pybox{volume.csv} file, the pressure trace file, and a file called
\pybox{volume-trace.yaml}, which contains the values that were set for each
attribute. The final step to conduct the simulations to calculate \(T_C\) and
the simulated ignition delay. This is done by the user by running the
\pybox{compare_to_sim()} function. This function takes two optional arguments,
\pybox{run_reactive} and \pybox{run_nonreactive}. These determine which
type(s) of simulation(s) should be conducted.

\begin{pythonbox}
cond_00_02.create_volume_trace()
cond_00_02.compare_to_sim(run_reactive=True, run_nonreactive=True)
\end{pythonbox}

UConnRCMPy is documented using standard Python docstrings for functions and
classes. The documentation is converted to HTML files by the Sphinx
documentation generator~\autocite{Brandl2016}. The format of the docstrings
conforms to the NumPy docstring format so that the autodoc module of Sphinx can
be used. The documentation is available on the web at
\url{https://bryanwweber.github.io/UConnRCMPy/}. UConnRCMPy also relies heavily
on functionality from the NumPy~\autocite{vanderWalt2011},
SciPy~\autocite{Jones2001}, and Matplotlib~\autocite{Hunter2007} Python
packages.

\section{Conclusions and Future Work}\label{conclusions-and-future-work}

UConnRCMPy provides a framework to enable consistent analysis of RCM data.
Because it is open source and extensible, UConnRCMPy can help to ensure that RCM
data in the community can be analyzed in a reproducible manner; in addition, it
can be easily modified and used for data in any format. In this sense,
UConnRCMPy can be used more generally to process any RCM experiments where the
ignition delay is the primary output.

Future plans for UConnRCMPy include the development of a robust test suite to
prevent regressions and document correct usage of the framework, as well as the
development of a plugin architecture to allow easy implementation of
user-defined analysis features. Other issues and directions are listed in the
Issue page of the GitHub repository
\url{https://github.com/bryanwweber/uconnrcmpy/issues/}.

\section{Acknowledgements}\label{acknowledgements}

This paper is based on material supported by the National Science
Foundation under Grant No. CBET-1402231.

\printbibliography

\end{document}